# Advanced Mass Calibration and Visualization for FT-ICR Mass Spectrometry Imaging

Donald F. Smith[1†,2], Andriy Kharchenko[1], Marco Konijnenburg[1], Ivo Klinkert[1], Ljiljana Paša-Tolić[2], Ron M.A. Heeren[1*]

1) FOM Institute AMOLF, Science Park 104, 1098 XG Amsterdam, The Netherlands
2) Environmental Molecular Sciences Laboratory, Pacific Northwest National Laboratory, Richland, WA 99352
   †: Current Address

* Address reprint requests to: Ron M.A. Heeren, Science Park 104, 1098 XG Amsterdam, The Netherlands, +31-20-754-7100, e-mail: heeren@amolf.nl

**Abstract**

Mass spectrometry imaging by Fourier transform ion cyclotron resonance (FT-ICR) yields hundreds of unique peaks, many of which cannot be resolved by lower performance mass spectrometers. The high mass accuracy and high mass resolving power allow confident identification of small molecules and lipids directly from biological tissue sections. Here, calibration strategies for FT-ICR MS imaging were investigated. Sub parts-per-million mass accuracy is demonstrated over an entire tissue section. Ion abundance fluctuations are corrected for by addition of total and relative ion abundances for a root-mean-square error of 0.158 ppm on 16,764 peaks. A new approach for visualization of FT-ICR MS imaging data at high resolution is presented. The "Mosaic Datacube" provides a flexible means to visualize the entire mass range at a mass spectral bin width of 0.001 Da. The high resolution Mosaic Datacube resolves spectral features not visible at lower bin widths, while retaining the high mass accuracy from the calibration methods discussed.

**Introduction**

Mass spectrometry (MS) imaging allows for the spatial localization of molecules from complex surfaces [1]. Though secondary ion mass spectrometry (SIMS) has long been used for high resolution spatial MS imaging [2], matrix assisted laser desorption ionization (MALDI) provides access to intact biological molecules such as lipids, peptides and proteins. MALDI MS imaging platforms are now commercially available, along with a suite of commercial and open-source software tools for image generation. Fourier transform ion cyclotron resonance mass spectrometry (FT-ICR MS)[3] offers the highest mass resolving power and mass accuracy for MS imaging experiments and has been used to spatially map neuro-peptides [4], lipids [5-7] and small molecules [8,9] from biological tissue sections. Further, orbital trapping FT-MS (i.e. the Orbitrap mass analyzer from Thermo Fisher Scientific)[10] is also seeing increased use for MS imaging of lipids [11-13], drugs [14] and peptides [15-17].

Both FT-ICR and orbital trapping (hereafter referred to collectively as FT-MS) require long time domain signals for ultimate mass resolving power and mass accuracy. This can result in large amounts of raw data, which must be processed and converted into a format amenable for visualization. In general, two approaches predominate; a "continuous" data format and "feature-based" (reduced and/or peak-picked) format.

The continuous format uses an average (or summed) mass spectrum, where ion signals can be selected from the mass spectrum for visualization.

This is the basis of the AMOLF-developed "Datacube" and Datacube Explorer visualization software, BioMap from Novartis [18], and FlexImaging from Bruker Daltonics.  However, it is often necessary to compromise mass spectral resolution (by binning the mass spectrum) in order to analyze the entire mass range at once.  This is due to the large amount of mass spectral information that must be read into computer memory (random access memory; RAM), especially when long time-domain transients are collected for high mass resolving power.  For FT-MS, mass spectral binning is undesirable as it negates the use of the high mass resolution analyzer.  The continuous format offers many advantages, which include visualization of the true mass spectral signal, easy user interaction with the mass spectrum and insurance that low intensity ions of interest are retained for image selection.  Disadvantages of the continuous format include the loss of low abundance spectral features due to averaging with noise/chemical background [6] and the aforementioned issues with large dataload at narrow mass bin widths.

Alternatively, the use of reduced data is advantageous as it can reduce the dataload significantly.  The MITICS software package reduces MS imaging data to XML format before further analysis and processing [19].  McDonnell et al have developed a workflow for automated feature detection from FT-ICR MS imaging datasets, where an average mass spectrum or a "base peak" mass spectrum is plotted and peaks are picked above a desired signal-to-noise ratio (S/N) and basepeak intensity threshold [6].  These features are then extracted from each pixel in the dataset, plotted as images and stored in this reduced format (which can be used for further statistical analysis). In addition, reduced data is stored by default for instruments from Thermo Scientific (e.g. LTQ-Orbitrap and LTQ-FT). The Spengler group has developed the MIRION software for generation of mass spectral images at any *m/z* bin width from centroid or profile raw data [16].

In addition to high mass resolving power, FT-ICR MS offers high mass accuracy which increases the specificity for identification of species detected in MS imaging experiments. In this context, mass calibration is essential for the ultimate chemical specificity of mass spectral images. Cornett et al have employed internal calibration (with respect to matrix related ions) for MALDI imaging of pharmaceuticals (and their metabolites) that resulted in under 1 part-per-million (ppm) standard deviation for olanzapine in rat liver [8]. Similarly, a lock mass calibration has been used for low ppm error Orbitrap FT-MS imaging of lipids [12] and peptides [16]. We have implemented an internal calibration of adjacent standards (INCAS) method that drastically improved mass accuracy for lipids detected by our custom-built FT-ICR MS [7]. The heterogeneity of biological tissue sections, as well as the shot-to-shot variability of MALDI, can introduce ion-number fluctuations that will influence the mass accuracy of FT-MS measurements. (For more information on ion-population dependent space-charge induced frequency shifts in FT-ICR MS, see the following and references therein [20-26].) Thus, calibration equations that account for ion number fluctuations should improve mass accuracy for FT-MS imaging experiments [23,27-30,25,31-33,26]. However, it should be noted that such calibration equations do not take into account other interactions in the ICR cell which can affect the measured frequency, such as image charge when ions approach the detection electrodes, instability of electric trapping fields or instability of the magnetic field.

Here, we compare four calibration methods for FT-ICR MS imaging and evaluate them in terms of root-mean-square (rms) mass accuracy, as well as analysis of ppm error histograms. Correction for total ion and relative ion abundance in the calibration equation improves measured mass accuracy and can be done so in an external manner. We illustrate ppm rms error for common lipids detected from a mouse brain tissue section, as well as for non-endogenous oligosaccharides that were spiked locally on the tissue surface. This high mass accuracy, coupled with the high mass resolving power of FT-ICR MS, allows very narrow bin sizes for ion selected images. A method for production of a continuous data format "mosaic Datacube" is described, which results in ultra-high mass resolution visualization of the entire mass range of interest. Ion selected images using a mass bin width of 0.001 Da are reported and the importance of full mass resolution is demonstrated. Finally, the mosaic Datacube is compared to results from a feature-based strategy that employs standard apex peak picking for automated feature extraction.

**Experimental**

*Materials and FT-ICR MS*

MALDI FT-ICR MS imaging experiments were performed on a 9.4 T solariX FT-ICR MS (Bruker Daltonics, Billerica, MA) in positive-ion mode. Transients of 2 mega-word were collected for an experimental mass resolving power ($m/\Delta m_{50\%}$) of 180,000 at $m/z$ 700. A mouse brain (female type 9 CFW-1, Harlan, Boxmeer, The Netherlands) was sectioned coronally to 12 µm on a cryo-microtome (Microm International, Walldorf, Germany). The section was coated with 20 mg/mL 2,5-dihydroxybenzoic acid (DHB; 1:1 methanol/water (0.2% trifuoroacetic acid)) using a Bruker ImagePrep. A 1 µL aliquot of oligosaccharide standards (Sigma-Aldrich, St. Louis, MO: maltotetraose, maltopentaose, maltohexaose, maltoheptaose; 555 µM each in 70/30 ethanol/water) was spotted on top of the tissue section post-matrix deposition (see **Supplementary Table 1**). The laser was operated at 1,000 Hz with 200 laser shots collected at each position with a stage raster size of 200 µm. An internal calibrant (ESI-L Electrospray Tuning Mix, Agilent Technoligies, Santa Clara, CA) was supplied by the ESI part of the dual ESI/MALDI source (similar to methods described previously [34,27]). The "End Plate Offset" voltage for the ESI source (as defined in solariXcontrol) was adjusted such that the intensity of the calibrant ions was similar to that of the lipids from the tissue section. In this way, 97% (2718/2777) of the spectra contained known ions that could be used for internal calibration. Electrospray instability caused the remaining 3% of spectra to be void of the ESI-L internal calibrants.

*Data Analysis*

Chameleon, an in-house developed workflow based data processing software built on the Microsoft .NET framework, was used to process the raw FT-ICR MS transients [35]. **Figure 1a** shows a standard Chameleon workflow for FT-ICR MS imaging. The transient is read into Chameleon, apodized and zero-filled (here, exponential apodization with two zero-fills), fast-Fourier transformed (FFTW library[36]), calibrated from frequency to mass space, apex peak picked (here, a threshold corresponding to S/N > 8 at *m/z* 700 was used) and the spectrum inserted into the "Datacube" structure developed at AMOLF. This process is repeated until the end of the dataset.

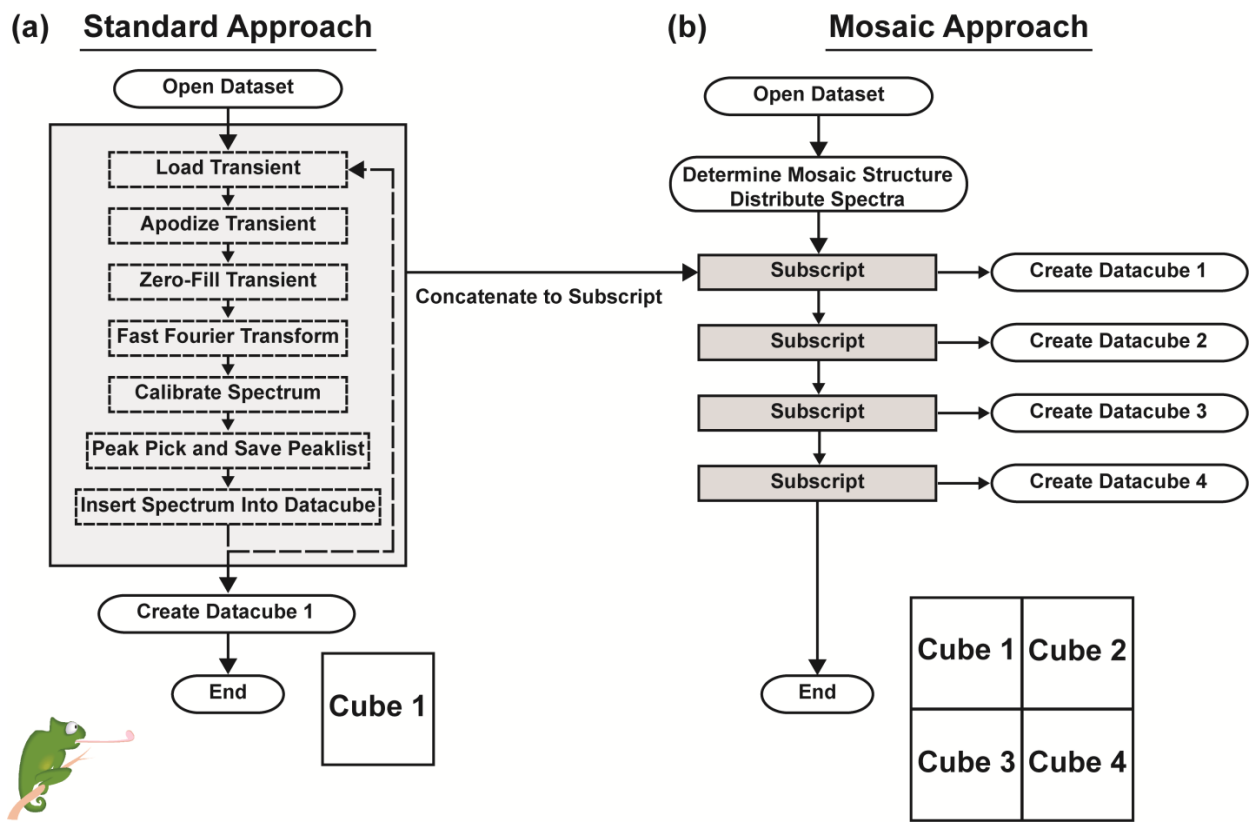

**Figure 1.** Workflows implemented in the Chameleon data processing software for FT-ICR MS imaging datasets. (a) Standard approach for creating a single Datacube, (b) Mosaic approach, where multiple spatially distributed Datacubes are created in a single analysis.

Four calibration methods were evaluated; external, lock mass, internal, and abundance corrected. Internal and external calibration used the calibration method described by Ledford (see Supplemental Material equation S1) [22]. Comparison of the calibration routine described by Ledford and that described by Francl [21] yielded similar mass measurement accuracy results, in accordance with previous studies (data not shown) [37]. Internal calibration was done within Chameleon. The mass spectra are first peak-picked to locate the internal calibrant peaks and then recalibrated with respect to these internal calibrants. Lock mass calibration was performed on *m/z* 622 from ESI-L in Matlab (MATLAB version 7.13.0.564 (64 bit), Mathworks, Natick, USA) using the "proportion" method described by Wenger (Supplemental Material equation S2) [38]. The multiple linear regression method described by Muddiman et al. was used to correct for total ion abundance and relative ion abundance fluctuations (Supplemental Material equation S3) [29,31,32]. This method adds two terms to a modified "Francl" calibration equation, which were calculated by multiple linear regression (Microsoft Excel) from 12 spectra that covered the range of total ion abundance observed throughout the MS imaging experiment. This calibration equation formally accounts for the total ion abundance in the ICR cell ($A_{Total}$), with the additional calibration coefficient, $\beta_2$. Similarly, the relative ion abundance of each species ($A_{Relative}$) is also formally accounted for, with the additional calibration coefficient $\beta_3$. Here, the additional calibration coefficients for total ion and relative ion abundances ($\beta_2$ and $\beta_3$) are calculated off-line and then entered into Chameleon. The total ion abundance ($A_{Total}$, the summation of all spectral intensities) is then calculated within Chameleon and each spectral point corrected for its relative abundance ($A_{Relative}$, the intensity/height of the point). It should be noted that Chameleon uses the calibrated mass spectrum for Datacube generation for the external, internal and abundance corrected methods. Thus, the summed spectrum visualized in the Datacube Explorer is very well calibrated.

In-house developed Matlab code was used for automated feature extraction. Here, peak lists from the abundance corrected calibration (apex baseline peak picking > S/N 5) are read into Matlab and aligned using the LIMPIC algorithm [39] (3 ppm alignment tolerance) to create a master peak list. Peaks within a window of ± 0.001 Da of one of the "master" peaks are considered the same and their intensities are written into an X-Y array used for image generation. Note that while apex peak picking is implemented here, further investigation of alternate peak-picking strategies (and peak shape models) for optimum mass accuracy will prove beneficial for feature based data analysis methods.

Lipids were identified by comparison to the LIPID MAPS database (LIPID Metabolites and Pathways Strategy; http://www.lipidmaps.org) using a mass threshold of 0.005 Da. Lipids with an odd number of carbons were not considered due to their low occurrence in higher order animals [40].

**Results and Discussion**

*Mass Calibration*

Careful mass calibration is necessary to insure accurate representation of selected ion images. Four calibration methods were compared to determine their performance and robustness for FT-ICR MS imaging. After calibration, the mass measurement accuracy was calculated for eight highly abundant diacylglycerophosphocholine lipids commonly observed in positive-ion mode

MALDI MS imaging (see **Supplementary Material Table 2**). Peaks within 5 ppm of the lipid exact mass were considered, which resulted in a total of 16,764 peaks. The comparison of the four calibration methods is shown in **Figure 2** and in **Table 1**, where the histograms shown in Fig. 2 were modeled with a normal fit. The center of the histograms in Fig. 3 can be interpreted as the accuracy of the measurements (the normal distribution mean, µ, in Table 1) whereas the width of the histogram is related to the precision of the measurements (the normal distribution standard deviation, σ, in Table 1).

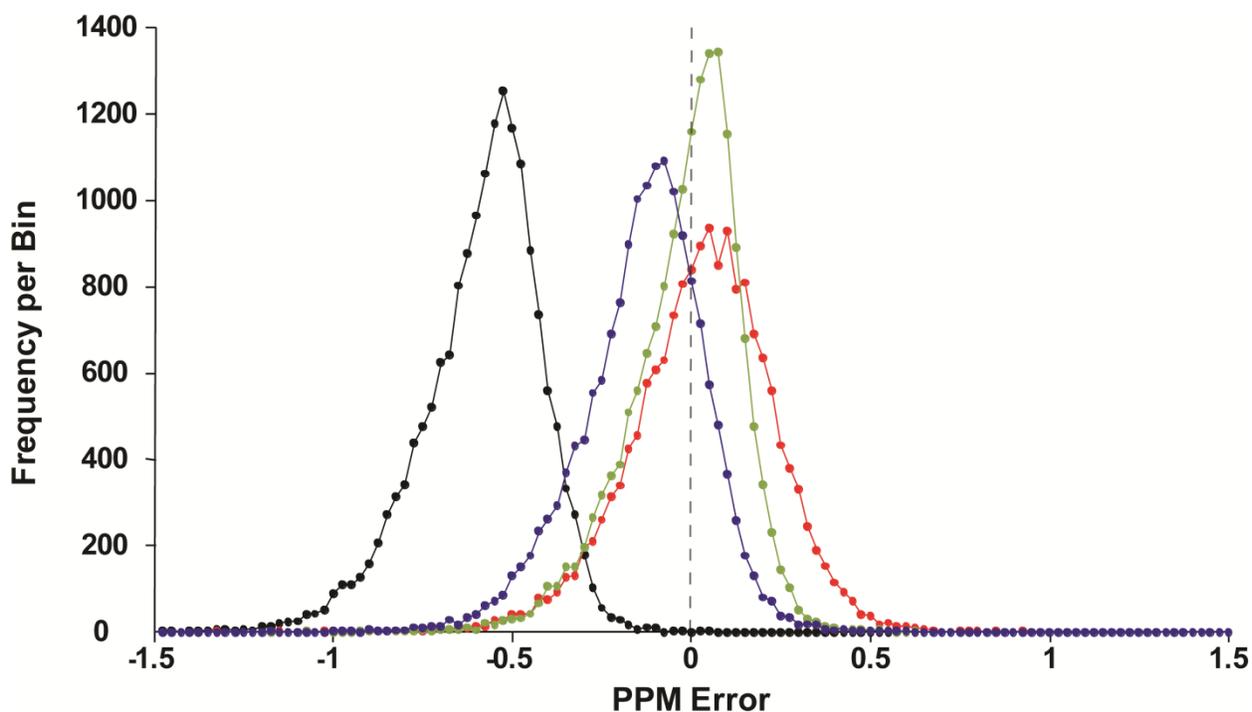

**Figure 2.** PPM error histograms for eight common diacylglycerophosphocholine lipids (bin size = 0.025 ppm). Black: external calibration, blue: lock mass ($m/z$ 622), red: internal calibration and green: abundance corrected. The lines have been added to guide the eye.

**Table 1.** RMS Error and Normal Distribution Figures of Merit for Common Diacylglycerophosphocholine Lipids from Mouse Brain

| Calibration Method | RMS Error (ppm) | Normal Distribution Mean ($\mu$, ppm) | Normal Distribution Standard Deviation ($\sigma$, ppm) |
|---|---|---|---|
| External | 0.609 | -0.586 | 0.168 |
| Lock Mass | 0.241 | -0.148 | 0.190 |
| Internal | 0.219 | 0.017 | 0.219 |
| Abundance Corrected | 0.158 | -0.018 | 0.157 |

With respect to mass accuracy, the external calibration performed the worst, with an rms error of 0.609 ppm. However, it is known that the quality of the external calibration is related to how well the total number of ions from the calibration spectrum is matched to the number of ions in the spectra to be calibrated. Thus, careful matching of ion populations can lead to much better external calibration. The lock mass performed well, with an rms error of 0.241 ppm. Internal calibration had an rms error similar to lock mass, albeit with a better normal distribution mean value. The abundance corrected calibration performed the best, with an rms error of 0.158 ppm. In terms of precision, external, lock mass and internal calibrations were all similar. This is expected, as these methods do not take any ion abundances into account. However, the abundance corrected calibration results in a narrower histogram distribution, which indicates that ion abundance differences are affecting the mass accuracy. However, it should be noted that low abundance chemical noise (below the mass spectral baseline) cannot be accounted for in the total ion abundance calculation and thus can still slightly affect mass accuracy.

A solution of oligosaccharides was spotted on the tissue for an additional assessment of the calibration methods. The rms mass measurement accuracy, as well as normal distribution mean and standard deviation for the oligosaccharide standards are shown in **Supplementary Material Table 2**. The results of all calibration methods have an rms error slightly higher than that of the endogenous lipids, which can be attributed to the higher mass of the oligosaccharides (Supplementary Material Table 1). The methods performed as follows (worst to best): external calibration, lock mass, abundance corrected, internal calibration. The slightly better performance of the internal calibration method versus the abundance corrected method can be attributed to the low total ion abundance in the area where the oligosaccharides were spotted (**Supplemental Figure 1**), which was not accounted for in the abundance corrected calibration. While this is an artifact of spotting the non-endogenous oligosaccharides to the surface, it illustrates the importance of selecting spectra with total ion abundance spanning a wide range for calculation of the abundance correction coefficients.

With the above performance characteristics in mind, implementation of the four calibration methods differs in complexity. External calibration is straight-forward, does not require re-calibration and is employed regularly for many applications. Lock mass for FT-MS imaging has been reported previously, where a ubiquitous matrix peak is typically used for calibration, making its implementation easy. Calibration on internal standards becomes more difficult. We have previously reported an INCAS method for a dedicated MALDI FT-ICR MS platform [7], while here the dual ESI/MALDI source allows easy injection of electrosprayed internal calibrants with the MALDI analytes from the tissue section. An attractive feature of the abundance corrected method is that it is implemented in an *external* manner. The correction terms can be calculated before or after the experiment and applied to the measured spectra during data analysis. This fact, coupled with its excellent performance, makes the abundance corrected calibration method highly attractive for FT-MS imaging experiments.

*Visualization of High Mass Resolution MS Imaging Data*

Proper visualization of high mass resolution FT-ICR MS imaging data becomes the priority after satisfactory calibration results are obtained. Methods using both continuous data and reduced data have been reported, but both ultimately contain a mass spectral binning step. As reported recently, bin widths for continuous format are generally limited by memory constraints (both random access and physical memory) dictated by the performance of the analysis computer[13]. To overcome these limitations, we have developed the "Mosaic Datacube", which uses smart data distribution and dynamic data selection to allow visualization of FT-ICR MS imaging datasets at 0.001 Da bin size. The creation of mosaic Datacubes within the Chameleon workflow is described in **Figure 1b**. The standard processing steps are concatenated into a subscript routine and the dataset is split into a user defined mosaic of individual Datacubes. In this way, the data is split into multiple smaller Datacubes, which alleviates RAM and disk storage space issues that arise from the need to store the entire dataset in one single file for visualization. Further, this architecture lends itself to parallel generation of Datacubes using multiple computers.

The challenge of storing high resolution MS imaging data in continuous format is illustrated in **Figure 3**. For Chameleon, the maximum Datacube size is 2 GB, which is the maximum allowable size of an array in the .NET framework (regardless of 32 or 64 bit versions). Thus, for a mass range of $m/z$ 300-1500, the minimum allowable $m/z$ bin size for this dataset is 0.0125. The minimum bin width is determined by memory limitations related to the number of pixels as well as the mass range of interest. Thus, for a smaller mass range, the $m/z$ bin width can be reduced. For a very narrow bin width of $m/z$ = 0.001, the maximum mass range for this experiment is only 100 $m/z$, as shown in **Fig. 3b.**

However, to visualize the entire dataset at such a narrow bin width requires that the data be distributed into smaller individual Datacubes. The desired number of Datacubes is manually entered into Chameleon, which then distributes the smaller cubes based on the number of pixels in the X and Y dimensions. Here, the dataset has been split into a 4x4 array of 16 individual cubes, as shown in **Fig. 3c**. The cube indicated in yellow represents a single Datacube (file) that spans the entire mass range. Now the entire dataset is not (and cannot) be read into memory on a typical desktop personal computer. Rather, a mass range of interest is defined in the Datacube Explorer and is read into memory on demand (indicated in red in Fig. 3c). Thus, the entire mass range can be visualized at 0.001 Da without compromising the high mass resolving power of the FT-ICR MS. Here, 0.001 Da was chosen as it corresponds well to the attainable mass resolving power of lipid species (m/z 700-900) with the chosen experimental parameters. At m/z 700, the mass resolving power is 180,000, which corresponds to a full width (at half maximum) of 0.004 Da. Thus, in order to visualize features separated by this value in the binned spectrum, it is necessary to bin at half of that value (i.e. 0.002 Da). Note that lower m/z species require a narrower bin size than 0.001 Da to not compromise the mass resolving power, as well as FT-MS experiments at higher mass resolving power (i.e. longer transients or higher magnetic fields), but at the cost of a larger dataload.

The mosaic Datacube approach is also applicable to data distribution in the mass space, where small mass range cubes (such as that in Fig. 2b) can be stored and dynamically read into memory in a similar manner. While the mosaic Datacube method does reduce the data size considerably (from 62.7 GB of raw data to 15.6 GB of mosaic Datacubes) it is still (physical) memory intensive. Further refinement of the data storage architecture is underway which will result in more compact mosaic Datacubes.

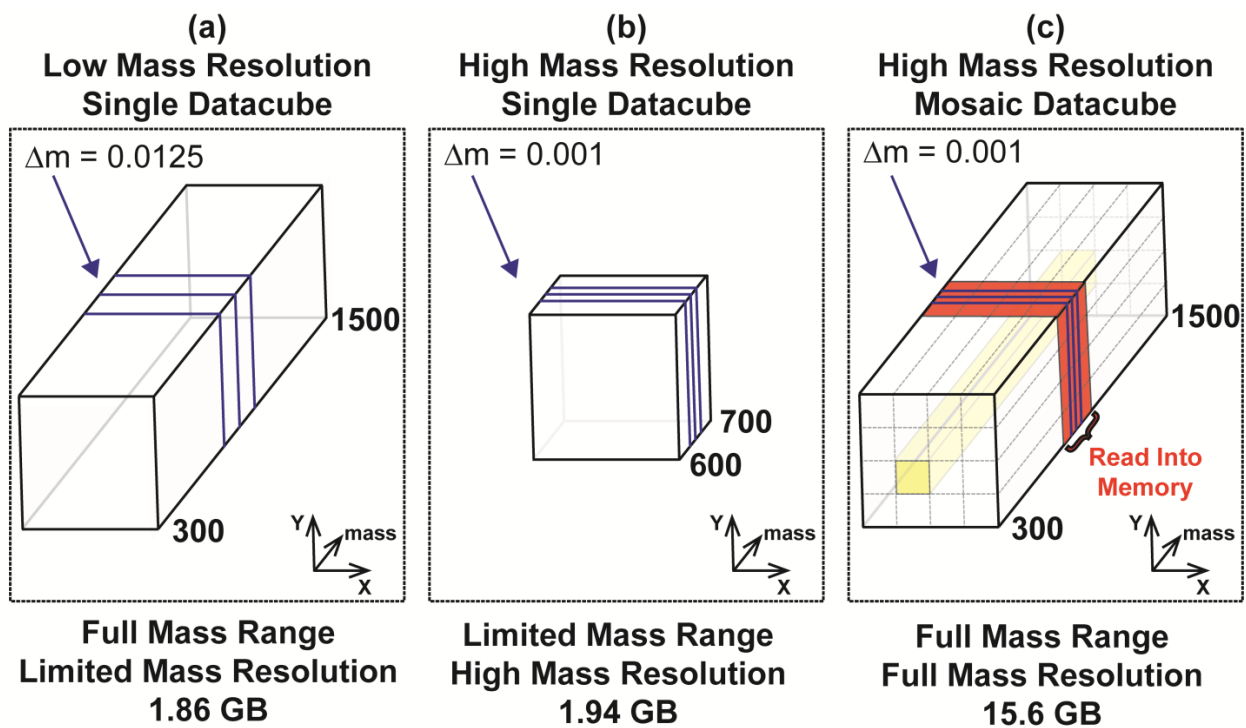

**Figure 3.** Comparison of different Datacube approaches. (a) full mass range with limited mass resolution, (b) high mass resolution with limited mass range, (c) high mass resolution and full mass range via the mosaic Datacube.

**Figure 4** shows a comparison of the summed spectra from a 0.0125 Da bin size (Fig. 4, top) and a 0.001 Da bin size mosaic Datacube (Fig. 4, bottom). The slight shift to higher mass of the upper spectrum is a result of the larger bin size, where the intensities within the bin are represented as the point at the end of the bin (e.g. for a bin from 784.00-784.0125, all intensities within the bin are represented by the point at 784.0125)[13]. The high mass resolving power of the FT-ICR MS is negated at a bin size of 0.0125 Da. While the ion selected ion images appear similar for the high intensity peaks, the smaller bin size of 0.001 Da reveals ions that are not resolved in the large bin size spectrum. The high mass resolution summed spectra ensure the highest selectivity for ion selected images, whereas ion selected images from the 0.0125 Da bin size are convolved with unresolved ions and thus uninterpretable. The high mass accuracy of FT-ICR MS allows the ions in Fig. 4 bottom to be assigned to phospholipids with high confidence. Assignment of lipids by exact mass has been shown previously. However, exact mass assignments would be erroneous if a summed mass spectrum with insufficient mass resolving power is used (such as Fig. 4a), as unresolved peaks would be assigned as the wrong species.

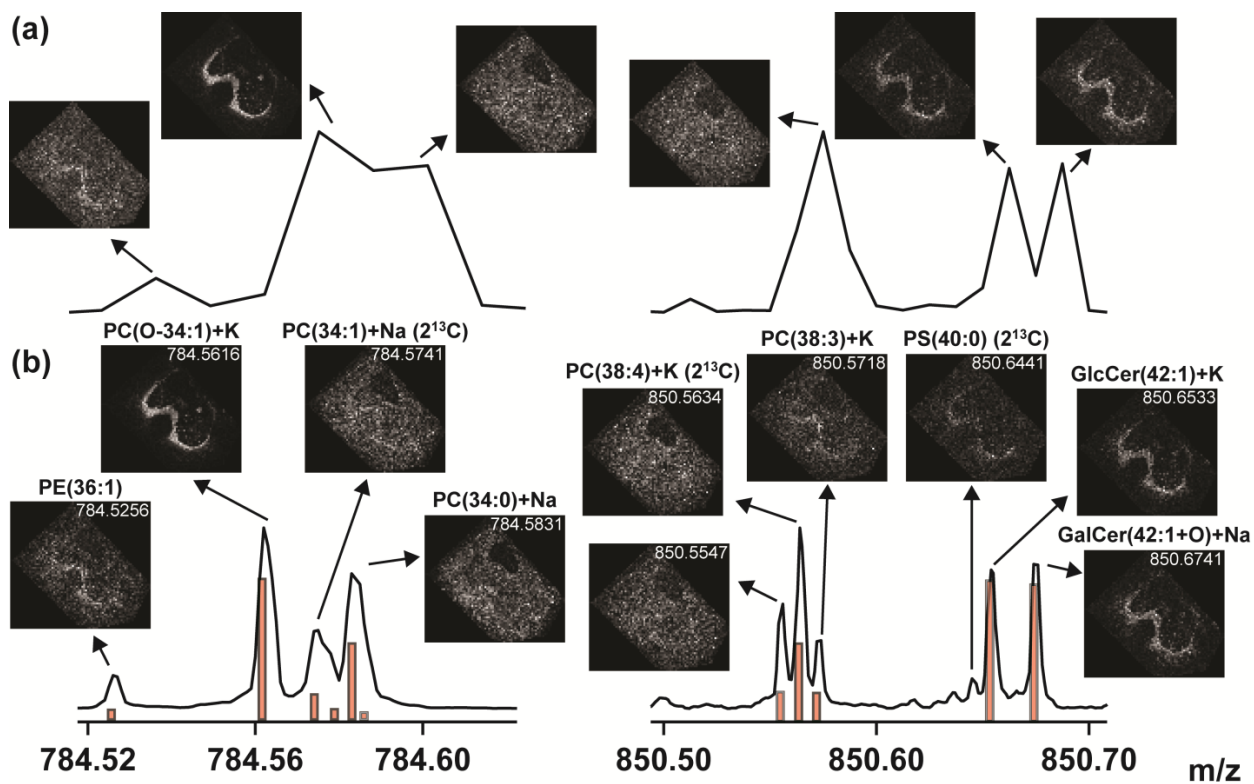

**Figure 4.** Comparison of mass bin sizes for Datacube generation (a) 0.0125 Da (entire dataset in one Datacube) and (b) 0.001 Da (Mosaic Datacube), where the results of peak picked feature based analysis are shown in red. Indicated mass values are the average across the entire dataset. A mass resolving power ($m/\Delta m_{50\%}$) of 100,000 is required to resolve the peaks at 850.5. Note that GlcCer and GalCer are isomers and cannot be distinguished by exact mass alone.

The data used to generate Fig. 4 was also peak picked (S/N > 5) and imported into in-house developed Matlab code for automated feature extraction. As noted previously[6,13], a large reduction in data size is realized by the use of feature based data. Here, the entire dataset of 62.7 GB is reduced to 7.3 MB (spectral representations, peak lists and image cube; 20 MB with images saved in jpeg format). The red bars in Fig. 4 bottom represent the features identified after LIMPIC alignment at 3 ppm. The intensity of the bars is the summed intensity of all peaks at that mass from the entire dataset. (Note, the intensity of the red bars and the Datacube spectrum are not directly correlated and have been adjusted for easy visualization.) The large peaks in the Datacube spectrum are well represented in the feature-based results. The shift to higher mass of the Datacube spectrum versus the feature-based results arises from the binning artifacts as described above. In fact, the difference in mass of the feature-based data and the apex of the peaks from the Datacube spectrum is ~0.001 Da (i.e. the bin size of the Datacube spectrum).

The feature-based approach yields peaks that are not visible in the Datacube spectrum, regardless of the small mass bin size of 0.001 Da. One example is observed as a "shoulder" of the peak at 784.5741 (PC(34:1)+Na ($2^{13}$C)). This clearly shows the advantage of the feature-based approach, since an even smaller mass bin size for the Datacube would be needed to resolve these two peaks, which would result in an even larger file size. However, at the S/N > 5 used here, the feature-based approach fails to pick low intensity peaks, such as 850.6441 (PS(40:0) ($2^{13}$C)). This can be remedied by global peak picking of the dataset, rather than on individual spectra as implemented here [6,13]. In combination, the continuous and feature-based approaches provide complimentary aspects for visualization of high mass resolution MS imaging data.

**Conclusions**

Calibration procedures for FT-ICR MS were investigated for high mass accuracy MS imaging. Correction for total ion abundance and relative ion abundance improves mass measurement accuracy for endogenous lipids. Such abundance correction should ensure high mass accuracy for samples with high regional molecular variability. The "mosaic Datacube" has been developed for high mass resolution visualization over the entire mass range at a mass bin size of 0.001 Da. The narrow bin width reveals features that are lost at higher bin sizes, but at the expense of a large dataload. Thus, future developments will focus on dataload reduction while maintaining the integrity of the high mass resolution data. Feature-based approaches relieve dataload concerns, but appropriate peak alignment and peak-picking thresholds must be chosen to ensure mass accuracy and low abundance peaks, respectively, are not lost. Integration of advanced calibration and visualization into the FT-ICR MS imaging data processing workflow allows routine use for low rms ppm error and high mass resolution visualization.

**Acknowledgments**

This work is part of the research programme of the Foundation for Fundamental Research on Matter (FOM), which is part of the Netherlands Organisation for Scientific Research (NWO). This publication was supported by the Dutch national program COMMIT. Portions of this research were supported by the American Reinvestment and Recovery Act of 2009 and the U.S. Department of Energy (DOE) Office of Biological and Environmental Research. The research described in this article was performed at the W. R. Wiley Environmental Molecular Sciences Laboratory (EMSL), a national scientific user facility sponsored by the Department of Energy's Office of Biological and Environmental Research and located at Pacific Northwest National Laboratory (PNNL). PNNL is operated by Battelle for the U.S. Department of Energy under Contract DE-AC05-76RLO 1830. The authors thank Si Wu for supplying the oligosaccharide standard, Nikola Tolić for help with the lock mass calibration and Julia Jungmann for assistance in debugging the Matlab code.

**Tables**

**Table 1.** RMS Error and Normal Distribution Figures of Merit for Common Diacylglycerophosphocholine Lipids from Mouse Brain

| Calibration Method | RMS Error (ppm) | Normal Distribution Mean ($\mu$, ppm) | Normal Distribution Standard Deviation ($\sigma$, ppm) |
|---|---|---|---|
| External | 0.609 | -0.586 | 0.168 |
| Lock Mass | 0.241 | -0.148 | 0.190 |
| Internal | 0.219 | 0.017 | 0.219 |
| Abundance Corrected | 0.158 | -0.018 | 0.157 |

**Figure Legends**

**Figure 1.** Workflows implemented in the Chameleon data processing software for FT-ICR MS imaging datasets. (a) Standard approach for creating a single Datacube, (b) Mosaic approach, where multiple spatially distributed Datacubes are created in a single analysis.

**Figure 2.** PPM error histograms for eight common diacylglycerophosphocholine lipids (bin size = 0.025 ppm). Black: external calibration, blue: lock mass ($m/z$ 622), red: internal calibration and green: abundance corrected. The lines have been added to guide the eye.

**Table 1.** RMS Error and Normal Distribution Figures of Merit for Common Diacylglycerophosphocholine Lipids from Mouse Brain

**Figure 3.** Comparison of different Datacube approaches. (a) full mass range with limited mass resolution, (b) high mass resolution with limited mass range, (c) high mass resolution and full mass range via the mosaic Datacube.

**Figure 4.** Comparison of mass bin sizes for Datacube generation (a) 0.0125 Da (entire dataset in one Datacube) and (b) 0.001 Da (Mosaic Datacube), where the results of peak picked feature based analysis are shown in red. Indicated mass values are the average across the entire dataset. Note that GlcCer and GalCer are isomers and cannot be distinguished by exact mass alone.

# Figures

# Figure 1.

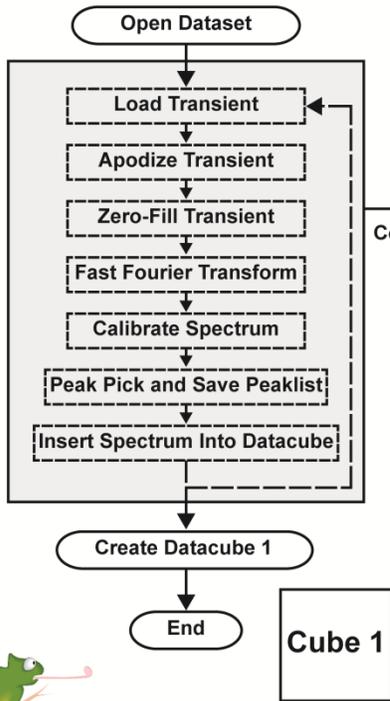
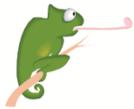
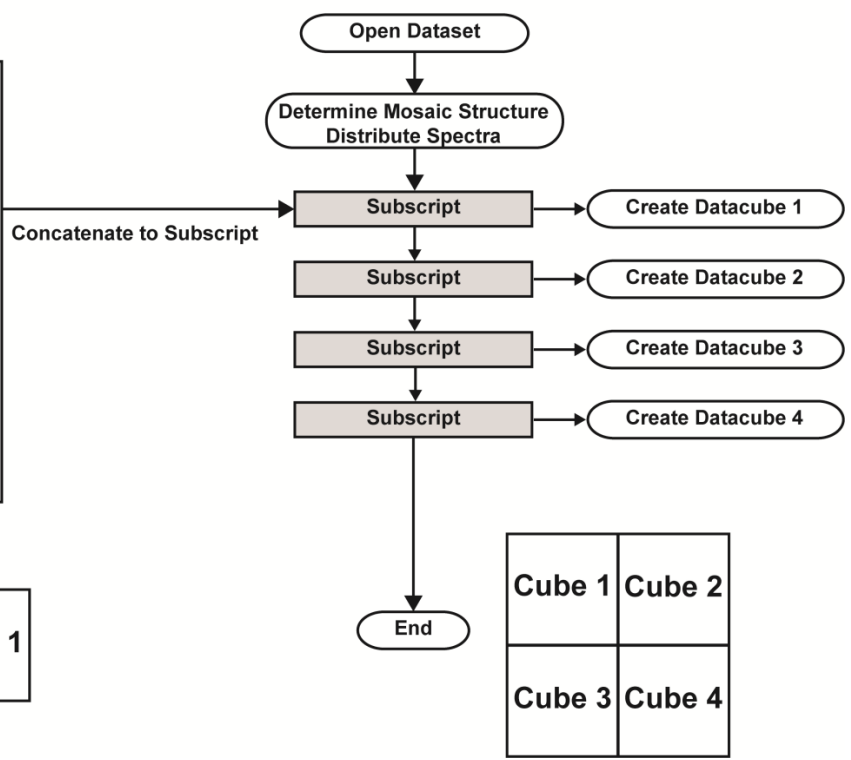

**Figure 2.**

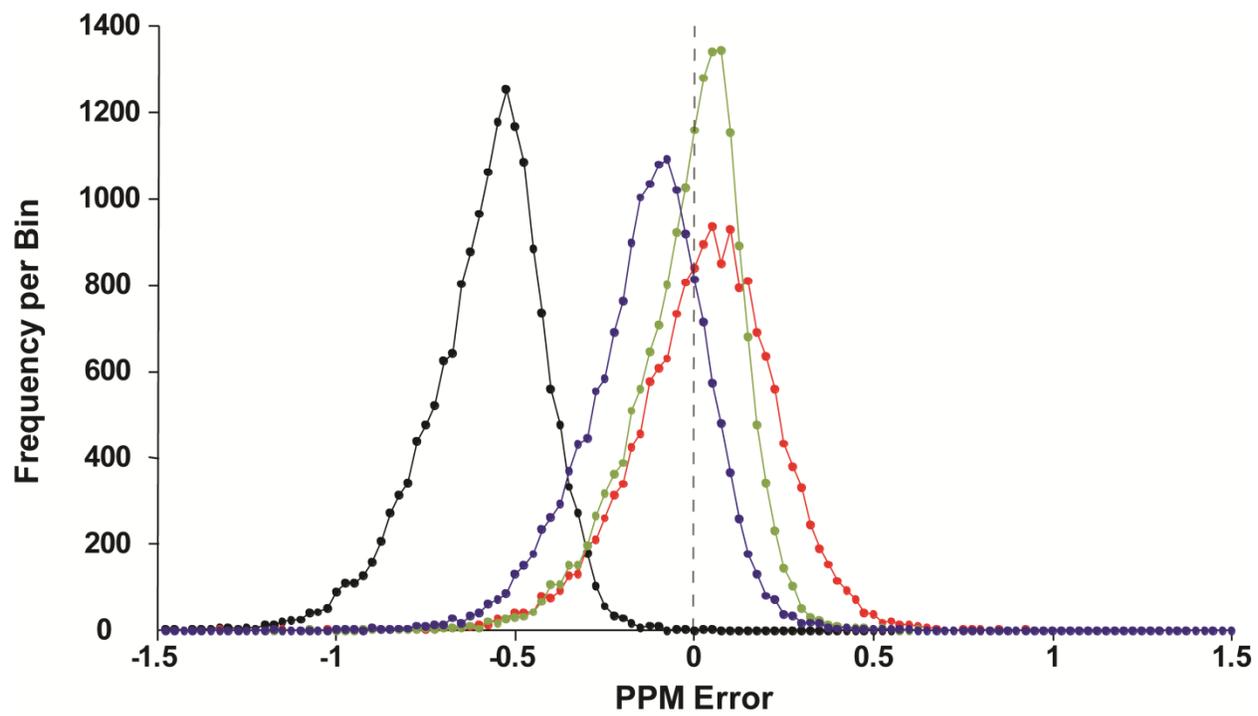

**Figure 3.**

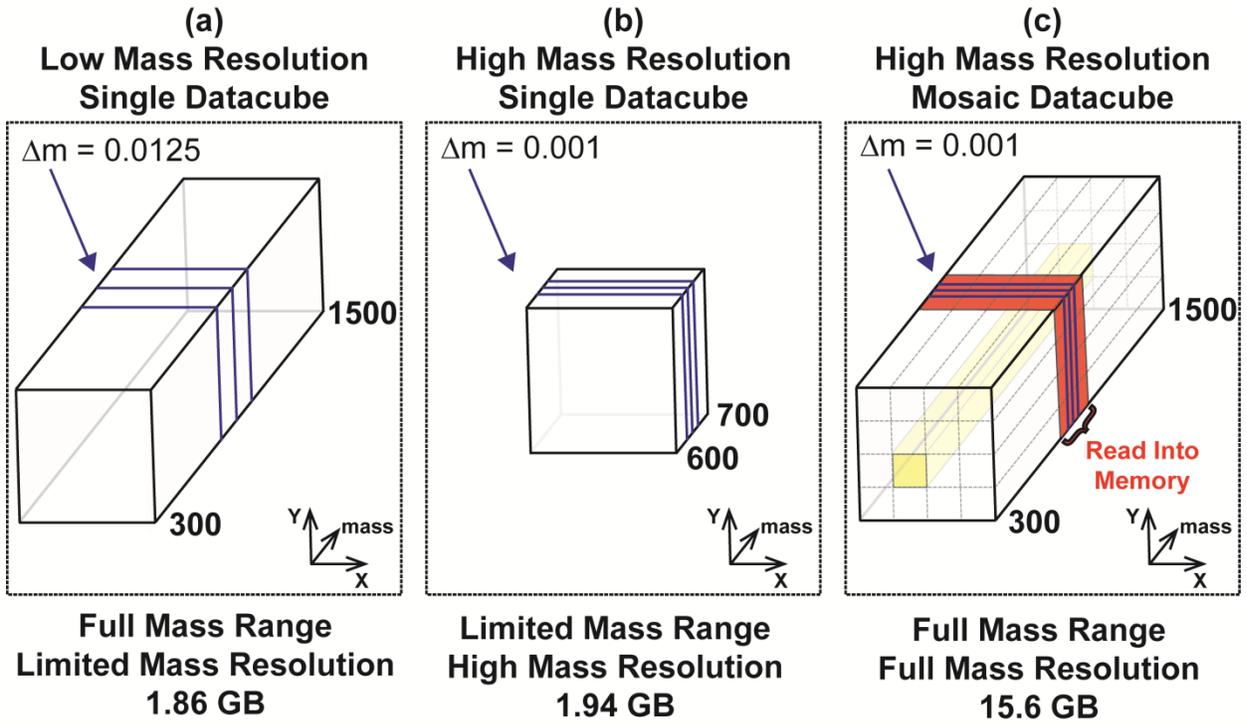

**Figure 4.**

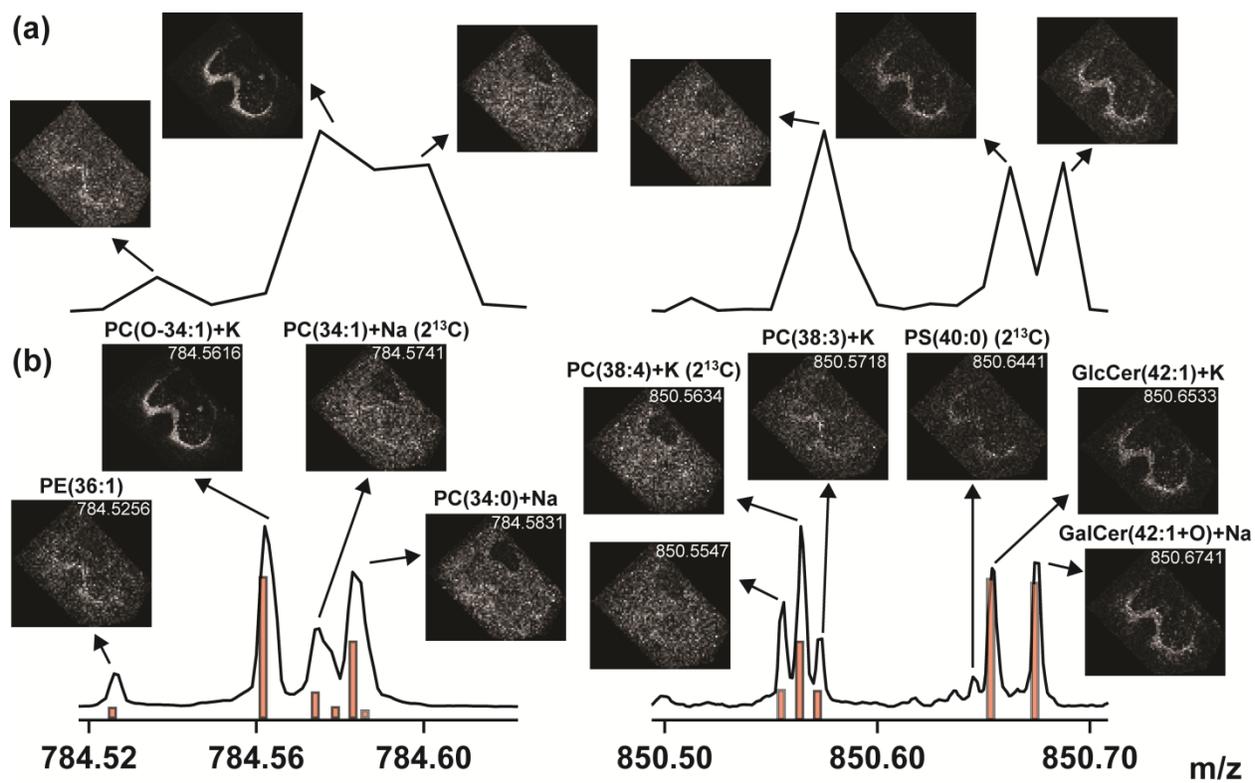

**<u>Supplementary Material</u>**

**Advanced Calibration and Visualization for FT-ICR Mass Spectrometry Imaging**

Journal of the American Society for Mass Spectrometry

Donald F. Smith[1†,2], Andrey Kharchenko[1], Marco Konijnenburg[1], Ivo Klinkert[1], Ljiljana Pasa-Tolic[2], Ron M.A. Heeren[1*]

1) FOM Institute AMOLF, Science Park 104, 1098 XG Amsterdam, The Netherlands
2) Environmental Molecular Sciences Laboratory, Pacific Northwest National Laboratory, Richland, WA 99352
   †: Current Address

* Address reprint requests to: Ron M.A. Heeren, Science Park 104, 1098 XG Amsterdam, The Netherlands, +31-20-754-7100, e-mail: heeren@amolf.nl

**Supplementary Table 1.** Identity and Exact Masses of the Oligosaccharides Spotted on the Mouse Brain Tissue

|  | Chemical Formula | $[M+H]^+$ | $[M+Na]^+$ | $[M+K]^+$ |
|---|---|---|---|---|
| Maltotetraose | $C_{24}H_{42}O_{21}$ | 667.2291 | 689.2111 | 705.1850 |
| Maltopentaose | $C_{30}H_{52}O_{26}$ | 829.2820 | 851.2639 | 867.2378 |
| Maltohexaose | $C_{36}H_{62}O_{31}$ | 991.3348 | 1013.3167 | 1029.2907 |
| Maltoheptaose | $C_{42}H_{72}O_{36}$ | 1153.3876 | 1175.3696 | 1191.3435 |

**Calibration Equations**

    <u>"Ledford" Calibration (External and Internal Calibration) [1]</u>

$$m/z = \frac{\beta_0}{f_{obs}} + \frac{\beta_1}{f_{obs}^2} \qquad (S1)$$

    <u>"Lock Mass" Calibration [2]</u>

$$m/z_{\text{recalibrated}} = m/z_{\text{original}} * \frac{m/z_{\text{recalibrated}}(\text{calibrant})}{m/z_{\text{original}}(\text{calibrant})} \qquad (S2)$$

    <u>"Abundance Corrected" Calibration (from multiple linear regression) [3-5]</u>

$$m/z = \frac{\beta_1}{[f_{obs} - (\beta_0 + \beta_2 A_{\text{Total}} + \beta_3 A_{\text{Relative}})]} \qquad (S3)$$

$f_{obs}$ = observed (measured) frequency

$\beta_0$ = magnetic field strength coefficient

$\beta_1$ = electric field coefficient

$\beta_2$ = total ion abundance (population) coefficient

$\beta_3$ = relative ion abundance coefficient

$A_{\text{Total}}$ = total ion abundance

$A_{\text{Relative}}$ = relative ion abundance of a given species

**Supplementary Table 2.** Endogenous Diacylglycerophosphocholine Lipids Used for Mass Measurement Accuracy Calculations

| Lipid Designation | Exact Mass |
|---|---|
| PC(32:0) | 734.5694 |
| PC(32:0)+Na | 756.5514 |
| PC(34:1) | 760.5851 |
| PC(32:0)+K | 772.5253 |
| PC(36:1) | 788.6164 |
| PC(34:1)+K | 798.5410 |
| PC(34:0)+K | 800.5566 |
| PC(38:6) | 844.5253 |

**Supplementary Table 3.** Root-Mean-Square ppm Errors and Normal Fit Figure of Merit Calculated for the Oligosaccharides Spotted on Mouse Brain Tissue

| Calibration Method | RMS Error (ppm) | Normal Distribution Mean (μ, ppm) | Normal Distribution Standard Deviation (σ, ppm) |
|---|---|---|---|
| External | 0.902 | -0.846 | 0.316 |
| Lock Mass | 0.477 | -0.350 | 0.325 |
| Abundance Corrected | 0.328 | -0.165 | 0.283 |
| Internal | 0.310 | -0.079 | 0.302 |

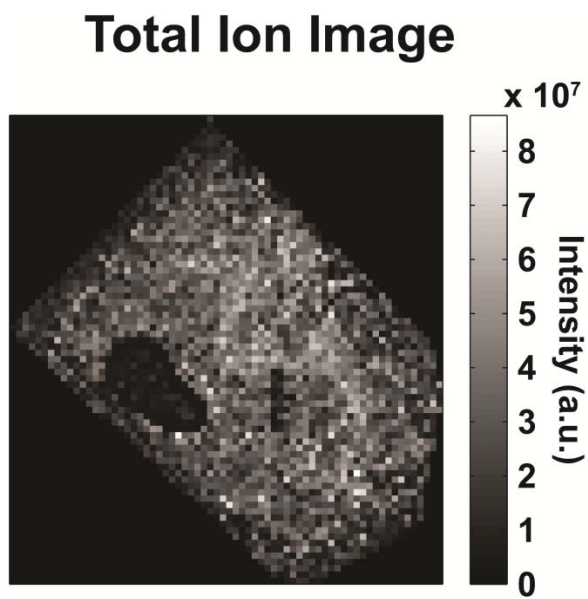

**Supplementary Figure 1.** Total ion image ($m/z$ 300-1500) as constructed from the feature-based data.